\documentstyle[multicol,prl,aps,epsf]{revtex}
\begin{document}

\title{A simple model for the formation of a complex organism
}  
\author{Barbara Drossel 
} 
\address{Theoretical Physics Group, University of
Manchester, Manchester M13 9PL, UK} 
\date{\today} 
\maketitle

\begin{abstract}
A simple model for the formation of a complex organism is introduced.
Individuals can communicate and specialize, leading to an increase in
productivity. If there are limits to the capacity of individuals to
communicate with other individuals, the individuals form groups that
interact with each other, leading to a complex organism that has
interacting units on all scales.
\end{abstract}
\pacs PPACS 89.90.+n, 05.70.Ln

\begin{multicols}{2}
During the past years, physicists have begun to study complex systems
like evolution, ecological systems, human civilization, and
economics. The models introduced for this purpose are usually composed
of units that interact according to a simple rule and produce a
complex large-scale behavior. While these models are still far from a
realistical description of the details of the systems under study,
they are capable of reproducing some of their essential features. Thus,
toy models for evolution can give rise to a power-law size
distribution of extinction events \cite{sne93,sole96,newman97,ama99},
models for ecological webs generate several trophic layers of
species \cite{cal98,can98}, models for urban development produce a
power-law size distribution of cities \cite{zan97,mar98,mak98}, and
models for stock exchange \cite{cal97} and company growth \cite{ama98}
show the scaling behavior characteristic for those systems. 

Most of these models focus on one organizational level, like the
formation of cities from interacting individuals, or the formation of
foodwebs from interacting species of variable abundance.  However, one
important characteristic of complex organisms like life on earth or
human civilization is that they have interactions between units of
various sizes. Thus, a biotope consists of several interacting
species, a species of interacting individuals, and individuals of
cells. Human civilization is structured into countries, which consist
of cities, which consist of smaller units like quarters and families,
etc.  

Phenomena of aggregation and clustering are widespread even in
inanimate nature. Thus, atoms may form stable clusters that preserve
their identity when aggregating to build a quasicrystal
\cite{jan97}. Aggregation in colloids and aerosols can be successfully
described using hierarchical models where clusters are repeatedly
joined to form larger clusters \cite{sor98}. Finally, on a cosmical
scale interactions between galaxies within galaxy clusters are
important for understanding the features of these clusters \cite{dub98}.

It is the purpose of this paper to introduce a model that
produces a complex organism with interacting units on various
organizational levels. Rather than trying to model a specific system
in some detail (which is done in \cite{vaa94}), we choose a model that
contains the essential ingredients in the simplest possible form. The
first of these ingredients is the capability of structural units to
interact or communicate and to specialize or differentiate, thereby
increasing a quantity that is called ``fitness'' in biology or
``utility'' in economics, and that will be called ``productivity'' in
this paper. This capability is acknowledged by biologists \cite{allee} as
well as by authors that adopt an evolutionary view of economics
\cite{sfproceedings,bor98} and societies \cite{parsons}.
Communication and differentiation alone, however, do not yet lead to
several levels of organization. They simply lead to one large group of
specialized individuals that has a high productivity.  We have to take
into account that the size of groups is restricted due to the limited
capacity of individuals to communicate and to travel. This restriction
naturally leads to hierarchical structures with several organizational
levels, as for example in the central place theory of human geography
\cite{geo}. The reason is that as soon as several groups exist, these
groups can communicate with each other {\it as groups}.  In a human
society, for instance, messengers are sent back and forth, roads are
built, and goods are traded that an independent individual could
neither produce nor make use of. This leads to a certain degree of
specialization among groups, and to a further increase in
productivity.  This argument can now be iterated by noticing that
groups have also a limited capacity to communicate or to interact. One
then obtains supergroups and groups of supergroups, etc.

Taking the above-mentioned basic ingredients into account, we define
our model in the following way: Let $P_1(n)$ be the productivity of a
group of $n$ individuals, and let $P_1(1)$ be negligible. For small
$n$, the productivity of a given member increases with the number of
partners with which it can communicate, the simplest analytical form
being a linear law with a parameter $g_1$ that is the productivity per
group member and communication partner.  For larger $n$, the cost of
communication must increase faster than this, and we may choose
\begin{equation}
P_1(n)=g_1n(n-1)-c_1n^2(n-1),
\label{p1}
\end{equation}
with $c_1 \ll g_1$. Such a law would, e.g., result if the
communication cost was proportional to the number of partners and to
the distance to each partner, and if the group extension grew linearly
in $n$. The index 1 indicates that the parameters are associated with
the first organizational level. The optimum group size, for which the
productivity per individual is largest, is $n=(g_1+c_1)/2c_1$. The
maximum possible group size, for which the productivity is not yet
negative, is $n=g_1/c_1$. The group size, above which a split into two
independent groups of size $n/2$ increases the productivity, is
$n=2(g_1+c_1)/3c_1$. For unequal splits, the productivity is smaller.

Now let us introduce interaction between groups. When the productivity
of a partner group is larger, the gain $g_2$ per group member due to interaction
with this partner will also be larger. Also, a larger group will put
more energy into communication. We may therefore write for the total
productivity of a ``supergroup'' consisting of $I$ interacting groups
\begin{equation}
P_2(n)= \sum_{i=1}^I P_1(n_i)+g_2 \sum_{i\neq j} n_iP_1(n_j) -c_2
I(I-1)\sum_{i=1}^I n_i.
\label{p2}
\end{equation}
The generalization to higher levels of organization is straightforward.

The parameters $g_2$ and $c_2$ must lie within certain limits for the
model to be meaningful. The productivity of a supergroup that consists
of only two interacting groups of ideal size (i.e., $I=2$), should be
of the same order of magnitude as the productivity of two independent
ideal groups, leading to $g_2 \simeq c_1/g_1$. The parameter $c_2$
should not be much larger than $g_1^2/c_1I$, if $P_2$ shall be
positive for supergroup sizes $I$. 

For the subsequent calculations we assume that the parameters are such
that group and supergroup sizes $n$, $I$, etc., are large, and $n-1$
and $I-1$ can be replaced by $n$ and $I$, making the analytical
expressions simpler. The first term on the right-hand side of
Eq.~(\ref{p2}) can also be neglected, since it is by a factor $1/I$
smaller than the second term.  Furthermore, since the productivity of
interacting groups of equal size near the optimum size is larger than
the productivity of interacting groups of different sizes (given the
same total number of individuals), we assume that all groups,
supergroups, etc., have approximately the same size. This is a kind of
mean-field approximation. For a fixed total number of individuals,
$N=\sum_{i=1}^I n_i$, the optimum number of groups is then obtained
from the condition $(\partial P_2/\partial I)_{N} =0$, leading to
$$-g_1g_2N^2I+2g_2c_1N^3-2c_2I^4=0.$$ For $N \ll g_1^4g_2/c_1^3c_2$,
the last term can be neglected, leading to $I=2c_1N/g_1$, implying
that the mean group size $\bar n=N/I$ is given by the optimum group
size. However, when $N$ and $I$ become larger, the last term becomes
important, and the mean group size increases. 

The optimum values of $N$ and $I$ for supergroups are found from the
conditions $(\partial P_2/\partial I)_{N} =0$ and $(\partial
(P_2/N)/\partial N)_{I} =0$, leading to $N=2g_1I/3c_1$ and $I=2
g_1^3g_2/27c_2c_1^2$. The size of a group within an optimum supergroup
is thus $4/3$ times the optimum size of an isolated group.  The number
of groups within a supergroup is of the same order as the number of
individuals $n$ within a group, if $c_2$ is of the order of $0.1 g_1$.
(Here, we used the previously derived condition that $g_2$ is of the
order $c_1/g_1$, or $1/n$.)

Next, let us estimate how the total productivity increases as fuction
of $N$. The productivity of a group of order $k$ has the form
$$P_k= g_k I_k^2I_{k-1} P_{k-1} - c_kI_{k-1} I_k^3$$ Here, $I_1$ is
the number of individuals in a group, $I_2$ the number of groups in a
supergroup, $I_3$ the number of supergroups in groups of supergroups,
etc.  The total productivity, divided by the total number of
individuals $N=I_1I_2\cdots I_k$, is therefore of the order
$g_1g_2\cdots g_k N^2/I_k$. All $I_k$ are of the same order, if $g_2$
to $g_k$ are of the order $c_1/g_1$, and the $c_k$ are of the order
$g_1^{2k-3}/c_1^{2k-4}$. Then, $P_k$ is of the order $g_1 N^2$, which
is comparable to the productivity of a single large group that has no
communication cost. The formation of a complex structure with groups
and interactions at all level is a very efficient way of keeping the
total communication cost low! This bears some similarity with the
formation of networks of rivers or blood vessels, where a hierarchical
structure optimizes drainage of water or supply with oxygen and
nutrients at a minimum cost for transport \cite{sin96,wes97}.

So far, we have discussed mainly systems where the productivity is
globally optimized. However, a realistic system cannot probe all
possible configurations in order to find the optimum, and furthermore
it is not likely to make rearrangements that require the breaking and
reconstruction of a large number of connections. As individuals are
added, the growth of a complex organism will follow pathways that
increase the productivity without going over larger ``barriers'',
i.e., through large rearrangements. It can be expected that there
exist a variety of different growth rules that, although they do not globally optimize productivity, lead to a complex
organism of high productivity. The following three examples for
explicit growth rules, and the numerical results (see Figs.1,2,3),
illustrate this: The parameters are for all simulations $g1=1,
c1=0.1,g2=0.4,c2=0.2,g3=0.05,c3=0.01$. They are chosen such that the
group and supergroup sizes are small in order to facilitate the
graphic representation of the growth process.  In the first simulation
(Fig.1), individuals are added to a group as long as this increases
the productivity of the group. Many isolated groups are formed
simultaneously. Then, groups start communicating with each other and
aggregate to supergroups. Supergroups grow until addition of further
groups does no more increase the productivity of the supergroup. Then,
supergroups start to aggregate. We also allow groups that are part of
a supergroup to grow further if this increases the productivity of the
supergroup. For the parameters used in the simulation, groups start to
aggregate at size 5 and grow during aggregation further up to size
7. When isolated supergroups reach the size 17, they start
communicating with each other to form groups of supergroups. The
simulation was stopped at this stage to allow for an easy graphical
representation of the result. If it was continued by adding more
individuals, several supergroups would form and aggregate to even
larger units, etc., thus producing an even higher hierarchy of
organizational levels.
\begin{figure}
\epsfysize=0.54\columnwidth{{\epsfbox{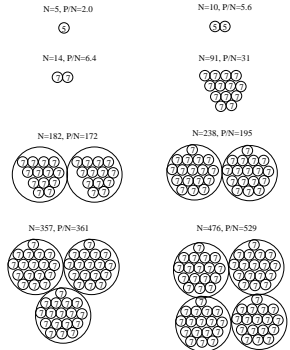}}}
\narrowtext{\caption{Growth of the system for the first set of
rules. The size $n$ of groups is indicated by the numbers in the small
circles. The large circles delimit supergroups. The parameters are $g1=1,
c1=0.1,g2=0.4,c2=0.2,g3=0.05,c3=0.01$\label{fig1} }}
\end{figure}

In the second simulation (Fig.2), we grow a complex organism by
adding individuals to it. We do not assume that other groups are
formed elsewhere that can later get in touch with each other. We allow
that individuals move to other groups, if this increases the
productivity. Thus, when a new individual is added, it may either join
one of the groups, or another individual that is part of a group may
join the newly added individual to open a new group, if this increases
the productivity. Once a new small group is started, further
individuals from larger groups can join it and increase the
productivity further. This happens for the parameters used in the simulations when the 6th individual is added. Similarly, we allow a group to split off a
supergroup to open a new supergroup (this can be seen in Fig.2 for $N=42$ and $N=168$), and we allow groups to move from
one supergroup to another, if this increases the productivity (see, e.g., the step from $N=42$ to $N=43$ in Fig.2). Thus,
all moves of single units (individuals, groups, etc.) are allowed that
increase the productivity.

In the third simulation (Fig.3), we also grow a single organism by repeatedly
adding individuals to it. As soon as this increases the productivity,
the initial group splits into two groups of equal size (this happens in Fig.3 for $N=6$). Further
individuals are added, until it becomes favorable to perform a
reconstruction into three groups of equal size, etc. As the number of
groups increases, the cluster of groups splits into two supergroups of
equal size, as soon as this increases the productivity (at $N=38$ in Fig.3), etc. Among the
three rules, the last rules produce the system with the highest
productivity, since it allows for the largest rearrangements.  

\begin{figure}
\epsfysize=1.42\columnwidth{{\epsfbox{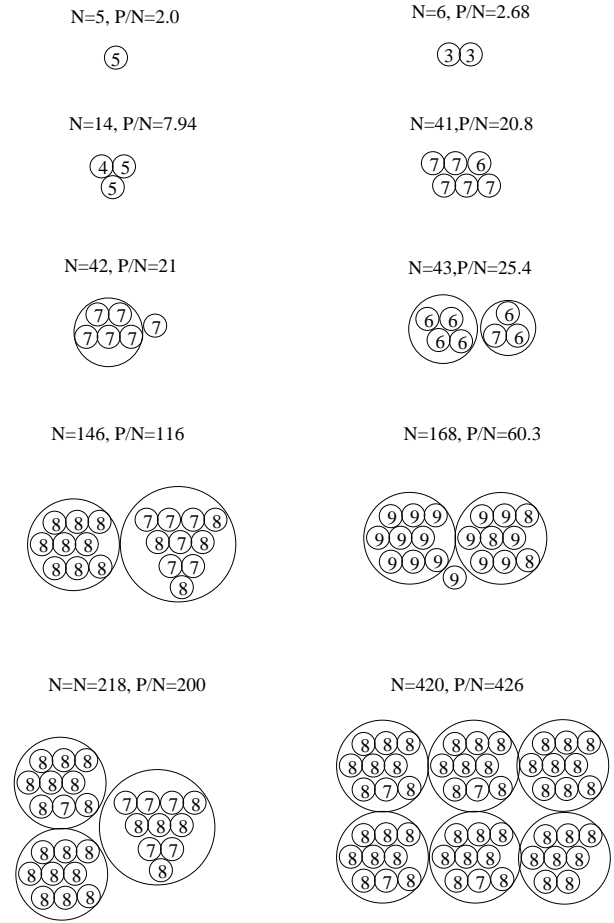}}}
\narrowtext{\caption{Growth of the system for the second set of
rules.\label{fig2} }}
\end{figure}

The three sets of rules together illustrate the many possible
dynamical pathways that can lead to the formation of a complex
organism. This indicates that the formation of a complex organism is a
generic phenomenon that can occur under fairly general conditions. The
main requirements are that communication increases the productivity,
and that the cost of communication exceeds the benefit if too many
units are involved. The specific expressions Eqs.~(\ref{p1}) and
(\ref{p2}) were chosen for their simplicity, however, many other forms
of the productivity function are possible.  

\begin{figure}
\epsfysize=1.35\columnwidth{{\epsfbox{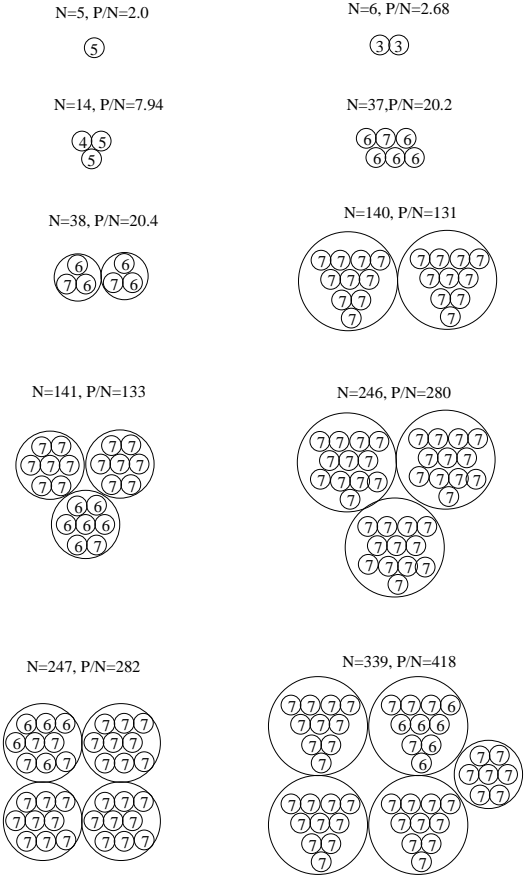}}}
\narrowtext{\caption{Growth of the system for the third set of
rules.\label{fig3} }}
\end{figure}
The model discussed in this paper assumes that all individuals and
groups are essentially equal. This is expressed, e.g., by the fact
that the parameters $c_k$ and $g_k$ are the same for all groups. One
can expect that a release of this restriction will still lead to the
formation of a complex organism. Also, the parameters of a complex
organism may change with time, which should not destroy the complexity
either.

The model presented here does not explicitely take into account that a
sufficiently large density of individuals is required for group
formation to occur. In the simulations, it was simply assumed that
enough individuals are around (or being born) for growth to
continue. It would be straightforward to include an explicit
dependence on distance in the communication cost, and to allow for
motion of individuals in space. These spatial degrees of freedom were
not considered in the present model to make the basic mechanism for
the increase in complexity more transparent.

In spite of its simplicity, this model agrees with recent results for
complex ecological webs \cite{can98}. Explicit models for the
interaction between several species show that the web becomes stable
if there is a sufficient number of weaks links. This condition is
naturally satisfied by the model presented in this paper, since each
individual interacts strongly only with the individuals in the same
group, but weakly with the rest of the system through links between
groups.

 \acknowledgements I thank H. Bokil, M.E.J. Newman, and J. Vaario for
comments and hints to the literature.
  This work was supported by
EPSRC Grant No.~GR/K79307.

\end{multicols} 
\end{document}